# Electrical Stability of $Cr_2O_3$/β-$Ga_2O_3$ and $NiO_x$/β-$Ga_2O_3$ Heterojunction Diodes

Yizheng Liu[1,a)], Haochen Wang[1], Carl Peterson[1], Chinmoy Nath Saha[1], Chris G. Van de Walle[1], and Sriram Krishnamoorthy[1,a)]

[1]*Materials Department, University of California Santa Barbara, Santa Barbara CA 93106, USA*

*a) Author(s) to whom correspondence should be addressed. Electronic mail: yizhengliu@ucsb.edu,, sriramkrishnamoorthy@ucsb.edu*

***Abstract***: This work reports the electrical characteristics comparison study between $Cr_2O_3$ and $NiO_x$ based heterojunction diodes (HJD) on halide vapor phase epitaxy (HVPE) grown β-$Ga_2O_3$ epitaxial layers. Both as-fabricated $Cr_2O_3$ and $NiO_x$ HJDs exhibited forward current density in a range of 130-150 A/cm$^2$ at 5 V with rectifying ratios >$10^{10}$ and a reverse leakage current density at $10^{-8}$ A/cm$^2$ at – 5 V. The differential specific on-resistance of $Cr_2O_3$ and $NiO_x$ HJDs was 12.01 mΩ•cm$^2$ and 12.05 mΩ•cm$^2$, respectively. Breakdown voltages of $Cr_2O_3$ HJDs ranged from 1.4-1.9 kV and 1.5-2.3 kV for $NiO_x$ HJDs. Theoretical band alignment between $Cr_2O_3$ and β-$Ga_2O_3$ was calculated from first principles. The ambient exposed $NiO_x$/HVPE β-$Ga_2O_3$ HJDs' forward current density degraded after 10 days while that of $Cr_2O_3$/HVPE β-$Ga_2O_3$ HJDs' remained nearly unchanged after the same amount of time. It was later confirmed that the ambient exposed sputtered $NiO_x$ sheet resistance ($R_{sh}$) degradation gave rise to the reduction of the forward current density of the $NiO_x$ based HJDs, and water ($H_2O$) was qualitatively determined to be the agent attributed to the forward conduction degradation by measuring the $R_{sh}$ of $NiO_x$-on-sapphire reference wafer after exposing it to different environments. The $Cr_2O_3$/HVPE β-$Ga_2O_3$ HJD also exhibited enhanced thermal stability compared to the $NiO_x$/β-$Ga_2O_3$ heterostructures at elevated temperatures. Interfacial nickel gallate ($Ga_2NiO_4$) phase formation expected from phase diagrams can explain the reduced thermal stability of $NiO_x$/β-$Ga_2O_3$ HJDs. This study indicates that $Cr_2O_3$ is a stable p-type oxide for the realization of robust multi-kV β-$Ga_2O_3$ HJDs.

# I. Introduction

In recent years of wide/ultra-wide bandgap (WBG/UWBG) semiconductor advancements, β-$Ga_2O_3$ offers promising potential for medium-voltage power applications (1-35 kV) in applications of grid transmission, renewable energy processing, and data centers for artificial intelligence (AI). The high critical electric field strength and availability of the shallow hydrogenic dopants[1] in epitaxial β-$Ga_2O_3$[2–5] can be leveraged to demonstrate power devices with much lower differential specific on-resistance while operating at a higher blocking voltage[6–8] compared to silicon carbide (SiC) and gallium nitride (GaN)[9,10]. Although reliable p-type doping with mobile holes in β-$Ga_2O_3$ is currently unavailable, p-type oxides, such as nickel oxide (NiO), have been integrated with β-$Ga_2O_3$ to form P-N heterojunctions that readily enable multi-kV junctions[11–16] with breakdown voltage >10 kV[17–19] and high electric field at 7.5 MV/cm[20].

Several materials issues in NiO still pose challenges in realizing reliable NiO/β-$Ga_2O_3$ heterojunction-based power devices. For example, instability of sputtered NiO under ambient exposure results in fluctuation of acceptor concentration ($N_A$) in this material, requiring a 15-minute nitrogen ($N_2$) anneal at 275 °C for stabilization[19,21–24]. Recently, it was also reported that the NiO/β-$Ga_2O_3$ heterojunction interface is susceptible to secondary nickel gallate ($Ga_2NiO_4$) phase formation[25,26], which affects the thermal stability of the junction characteristics. Stable p-type oxide/β-$Ga_2O_3$ heterojunction diodes with no interfacial phase formation are therefore necessary for reliable applications in β-$Ga_2O_3$-based power electronics.

Chromium oxide ($Cr_2O_3$), a p-type oxide[27] with a bandgap of ~3.6 eV[28,29], has recently been explored to form P-N heterojunctions[30–32] with β-$Ga_2O_3$ as an alternative to NiO-based junctions, exhibiting enhanced thermal stability under high temperature environment at 600 °C[32].

In this work, we compared the electrical characteristics of as-fabricated $Cr_2O_3$/HVPE β-$Ga_2O_3$ and $NiO_x$/HVPE β-$Ga_2O_3$ HJDs. Theoretical band alignment between $Cr_2O_3$ and β-$Ga_2O_3$ is reported by first-principles calculations. Forward current density of HJDs and oxide sheet resistance variation under ambient exposure were also tracked to compare the material stability. Water vapor ($H_2O$) in the air was qualitatively found to be the cause for degradation of the sheet resistance in sputtered $NiO_x$ exposed to ambient conditions. Temperature-dependent J-V measurements confirmed the enhanced thermal stability of $Cr_2O_3$-based HJDs at elevated temperatures. The enhanced thermal stability of $Cr_2O_3$/ β-$Ga_2O_3$ interface is further corroborated from the absence of a stable ternary phase in the Cr-Ga-O system compared to the $Ga_2NiO_4$ phase formation at the NiO/β-$Ga_2O_3$ heterojunction interface by comparing their theoretical phase diagrams.

# II. Device Fabrication

The $Cr_2O_3$/HVPE β-$Ga_2O_3$ heterojunction diode (HJD) fabrication began with a backside Ti/Au (50/150 nm) Ohmic metallization on $n^+$ β-$Ga_2O_3$ bulk substrate using e-beam evaporation followed by a 60-seconds rapid thermal annealing (RTA) at 470 °C in $N_2$. Following the backside Ohmic contact formation, a photoresist lift-off mask was patterned using optical lithography after standard solvent clean (acetone/isopropanol/de-ionized water). Then, a ~20-nm $p^-$ $Cr_2O_3$ layer ($R_{sheet}$ ~1 MΩ/□) was directly deposited on the HVPE β-$Ga_2O_3$ via reactive radio frequency (RF) magnetron sputtering by using a 99.95%-pure metallic chromium (Cr) target under an oxygen-deficient condition (Ar/$O_2$~22/10 sccm) at a 3.7 mTorr chamber pressure and 150 W RF power. To enhance the Ohmic contact quality to $Cr_2O_3$, a ~20-nm $p^+$ $Cr_2O_3$ contact layer ($R_{sheet}$ ~600 kΩ/□) was sputtered under an oxygen-rich condition (Ar/$O_2$~8/10 sccm) at a 3 mTorr chamber pressure and the same RF power. The atomic percentage of Cr (30.6%) and O (69.4%) were confirmed by using energy dispersive spectroscopy (EDS) on reference silicon monitor wafer, closely matching the stoichiometry of $Cr_2O_3$ reported[30] earlier. Immediately following the sputter deposition, a Ni/Au/Ni (50/50/200 nm) metal stack was deposited via e-beam evaporation to serve as an Ohmic contact to $Cr_2O_3$ and a hard mask for subsequent plasma dry etching[20,33]. The $Cr_2O_3$/Ohmic metal stack was later lifted off in a heated N-methyl pyrrolidone (NMP) solution. To mitigate the electric field crowding, the diodes were dry-etched ~1.2 μm below the $Cr_2O_3$/β-$Ga_2O_3$ heterojunction interface under inductively coupled $BCl_3$ plasma at 200 W, as shown in **Fig. 1(a)**. This self-aligned plasma-etch served as edge termination for the device. Similarly, the $NiO_x$/HVPE β-$Ga_2O_3$ HJDs, as shown in **Fig. 1(b)**, were also fabricated on the same wafer by using a sputtering condition of Ar/$O_2$~22/10 sccm at a 3.7 mTorr chamber pressure and 150 W RF power for ~20-nm of $p^-$ $NiO_x$ ($R_{sheet}$ ~215 kΩ/□) and Ar/$O_2$~8/10 sccm at a 3 mTorr chamber pressure and 150 W RF power for ~20-nm of $p^+$ $NiO_x$ ($R_{sheet}$ ~ 7–9 kΩ/□) by using the identical fabrication process[20,33]. The $NiO_x$ layers were sputtered from a 99.999% purity nickel (Ni) metallic target.

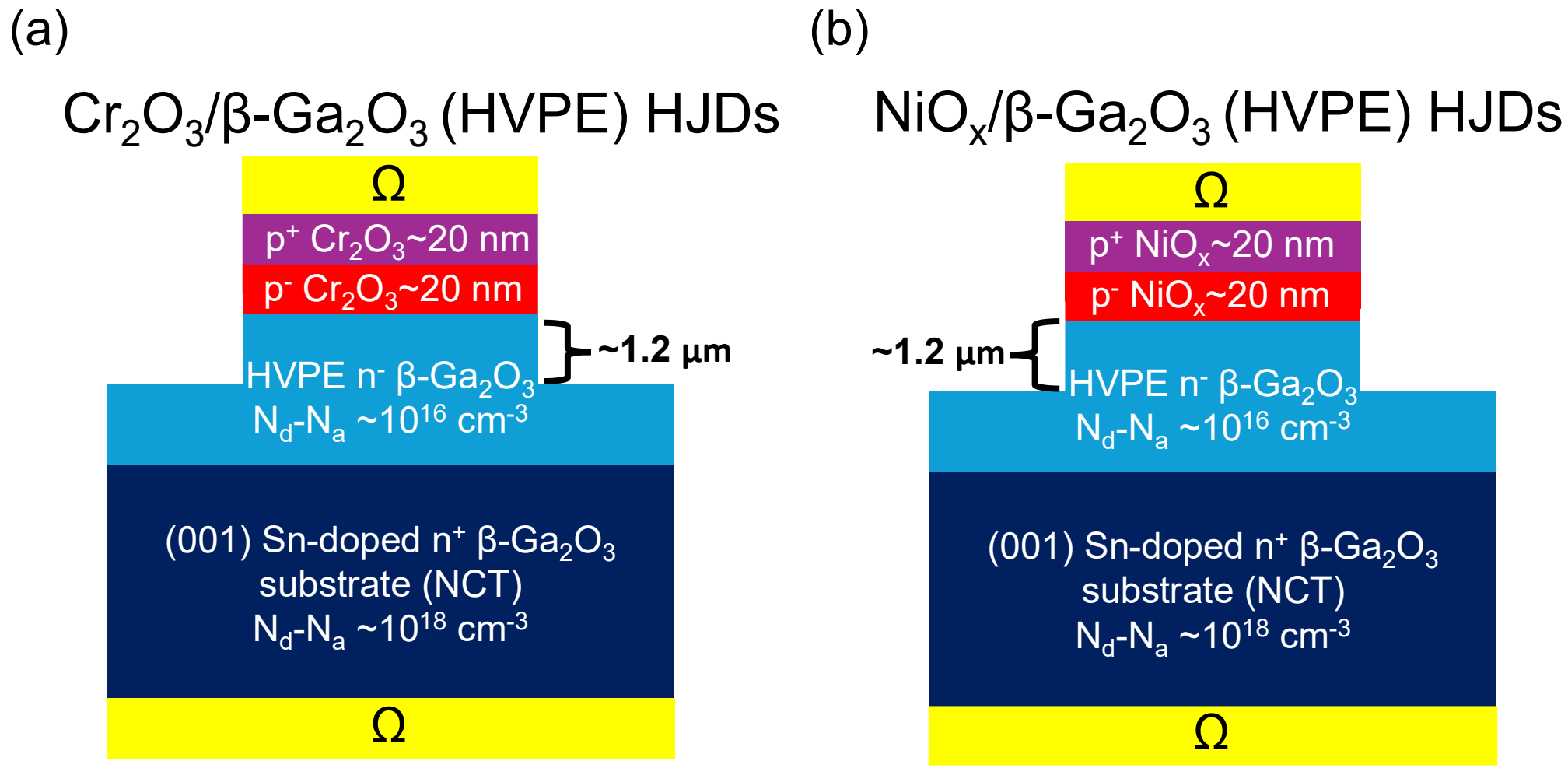

FIG.1. (a) $Cr_2O_3$/HVPE β-$Ga_2O_3$ heterojunction diode with ~1.2 μm plasma-etch edge termination schematics. (b) $NiO_x$/HVPE β-$Ga_2O_3$ heterojunction diode with ~1.2 μm plasma-etch edge termination schematics.

## III. $Cr_2O_3$/β-$Ga_2O_3$ Band Alignment Calculation

We use the electronic transition level of hydrogen to align the band structures.[34] We perform first-principles calculations to locate the (+/-) transition level of interstitial hydrogen in β-$Ga_2O_3$ and corundum $Cr_2O_3$. Our calculations are based on the generalized Kohn-Sham theory[35] using the HSE hybrid functional[36] implemented with projector augmented wave (PAW) potentials[37,38] in the VASP code.[39] In the HSE functional, the mixing parameters are fitted to reproduce experimental band gaps and lattice constants: 34% for β-$Ga_2O_3$ and 18% for corundum $Cr_2O_3$. With these parameters, we obtain for β-$Ga_2O_3$ an indirect band gap of 4.81 eV and lattice parameters $a$ =

12.25 Å, $b = 3.04$ Å and $c = 5.81$ Å. For corundum $Cr_2O_3$ we obtain an indirect band gap of 3.45 eV and lattice parameters $a = 4.97$ Å and $c = 13.65$ Å.

We employed 120-atom supercells; for β-$Ga_2O_3$ this corresponds to a 1×3×2 multiple of the 20-atom monoclinic conventional cell, and for corundum $Cr_2O_3$ to a 2×2×1 multiple of the 30-atom hexagonal conventional cell. Calculations are performed using a 2×2×2 k-point grid and a plane-wave cutoff energy of 500 eV. Spin polarization is included. For an interstitial hydrogen impurity in charge state $q$ ($\pm 1$), the formation energy is given by[40]

$$E^f\left(\mathrm{H}_\mathrm{i}^q\right) = E_t\left(\mathrm{H}_\mathrm{i}^q\right) - E_t(\mathrm{bulk}) - \mu_\mathrm{H} + qE_\mathrm{F} + \Delta^q, \tag{1}$$

where $E^f\left(\mathrm{H}_\mathrm{i}^q\right)$ is the formation energy of the interstitial hydrogen in charge state $q$, $E_t\left(\mathrm{H}_\mathrm{i}^q\right)$ and $E_t(\mathrm{bulk})$ are the total energies of the defective and pristine supercells, and $\mu_\mathrm{H}$ is the chemical potential of hydrogen. The term $qE_\mathrm{F}$ accounts for the exchange of electrons with the reservoir, where the Fermi level $E_\mathrm{F}$ is referenced to the VBM. $\Delta^q$ represents the finite-size correction for charged supercells, following established schemes.[41,42]

Due to the monoclinic crystal structure of β-$Ga_2O_3$, multiple configurations are possible for $\mathrm{H}_\mathrm{i}^+$ and $\mathrm{H}_\mathrm{i}^-$. In the lowest energy structures, $\mathrm{H}_\mathrm{i}^+$ bonds to a lone pair of three-fold coordinated oxygen atom whereas $\mathrm{H}_\mathrm{i}^-$ relaxes to a site near two Ga atoms within the mirror plane perpendicular to the $b$ axis. These configurations yield a (+/–) transition level 0.04 eV above the CBM as shown in **Fig. 2**, in agreement with previous work.[43] In corundum $Cr_2O_3$, both $\mathrm{H}_\mathrm{i}^+$ and $\mathrm{H}_\mathrm{i}^-$ bond to a four-fold coordinated oxygen atom. In the donor state ($\mathrm{H}_\mathrm{i}^+$), the four Cr-O bonds elongate by 3%, 4%, 6% and 10%. In the acceptor state $\mathrm{H}_\mathrm{i}^-$, three Cr-O bonds elongate by 0.5%, 15%, 20% while the fourth shortens by 1%. The resulting (+/–) level is located 0.44 eV below the CBM. In **Fig. 2**, we align the transition level (+/-) of interstitial hydrogen in $\beta$-$Ga_2O_3$ and corundum $Cr_2O_3$, resulting in a type-II band alignment for the heterojunction, with valence- and conduction-band offsets of $\Delta E_\mathrm{v}$=1.85 eV and $\Delta E_\mathrm{c}$=0.48 eV.

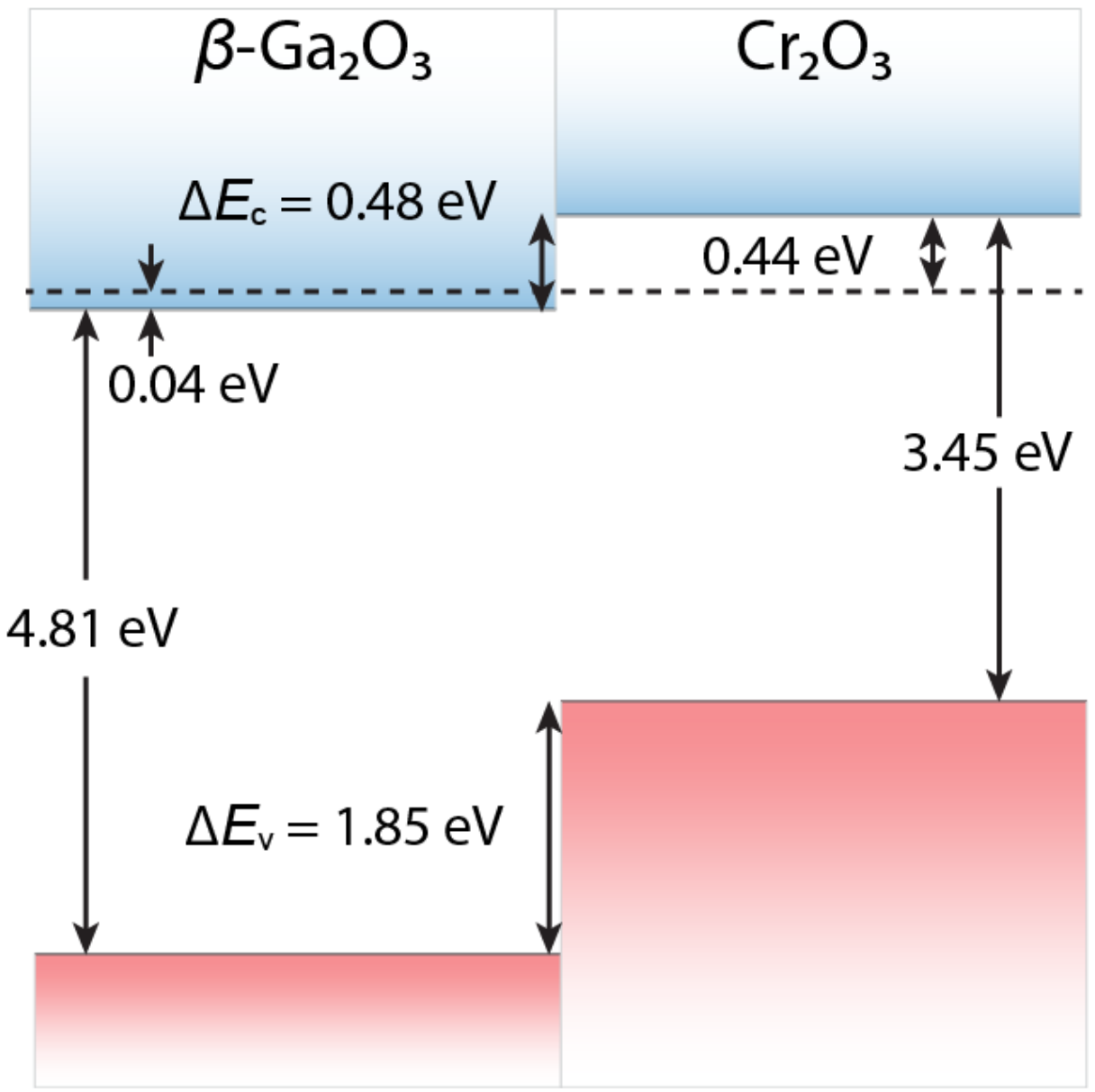

FIG.2. Band alignment between β-$Ga_2O_3$ and corundum $Cr_2O_3$. The valence bands are shown in red and the conduction bands in blue. The dashed line indicates the (+/−) transition level of interstitial hydrogen.

# IV. β-$Ga_2O_3$ HJDs Electrical Characteristics Analysis

## A. High-Voltage Capacitance-Voltage (C-V) Analysis

The punch-through electric field profile of $Cr_2O_3$/β-$Ga_2O_3$ HJDs beyond ~600 V was confirmed via high-voltage capacitance–voltage measurements at 1 MHz, as shown in **Fig. 3(a)**. The HVPE layer was fully depleted above 600 V reverse bias, where measured capacitance remained flat and no longer changed as a function of applied reverse bias. Moreover, the built-in potential ($V_{bi}$) of the HVPE HJD was extracted to be ~2 V from the $1/C^2$ vs voltage characteristics shown in Fig. 3(a) inset, closely matching the $V_{bi}$ extracted from $NiO_x$/β-$Ga_2O_3$ HJDs from our early reports[20,33]. The average HVPE drift layer's apparent charge density was extracted to be $9.5 \times 10^{15}$ $cm^{-3}$ with a corresponding thickness of ~7.48 µm using a relative permittivity of 12.4[44]for (001) β-$Ga_2O_3$, as shown in **Fig. 3(b)**.

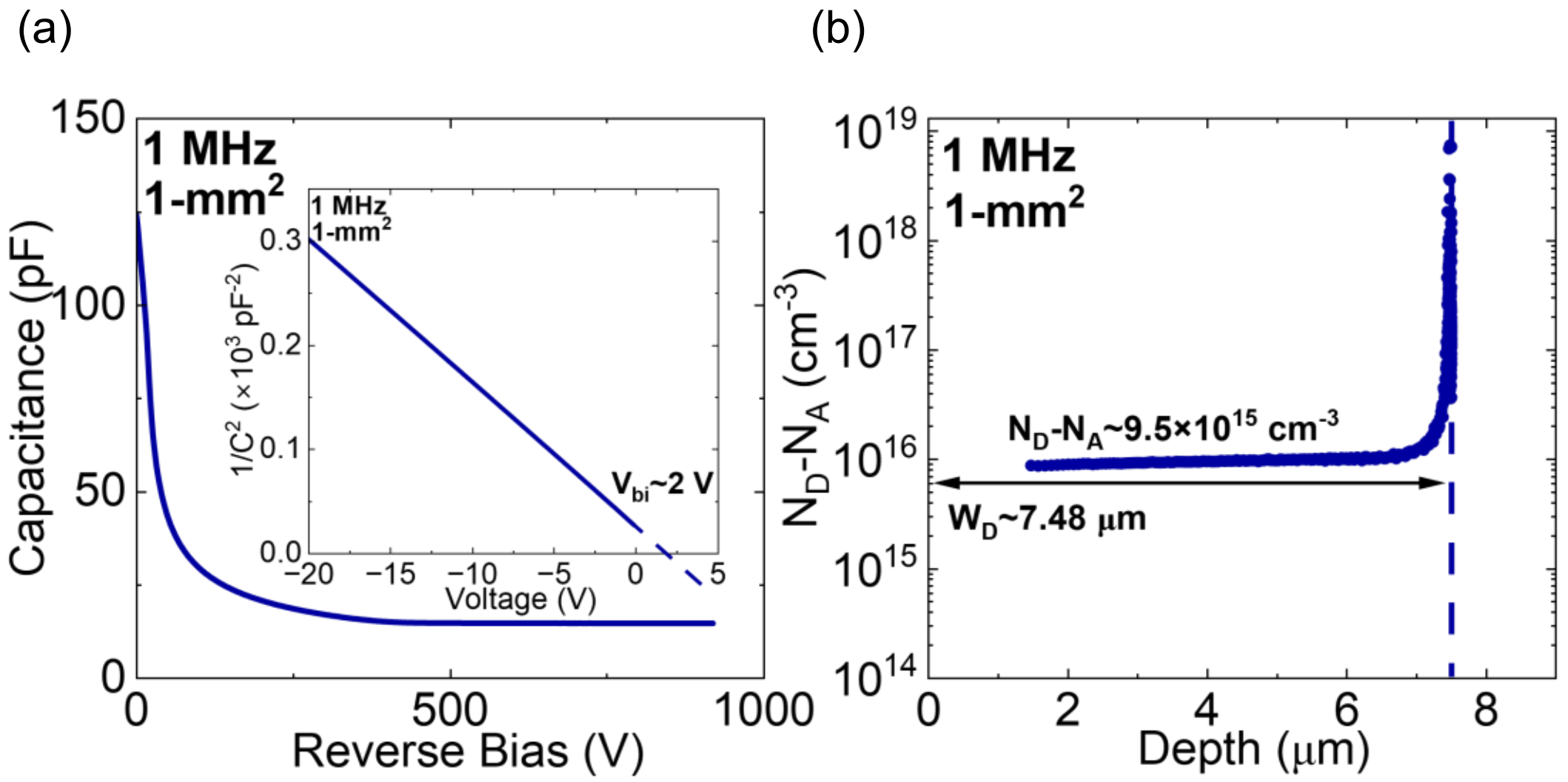

FIG.3. (a) High-voltage C–V characteristics on 1-mm$^2$ area pad at 1 MHz with inset showing the built-in potential of $Cr_2O_3$/HVPE β-$Ga_2O_3$ HJD. (b) Extracted apparent charge density vs depth profile of the plasma-etched $Cr_2O_3$/HVPE β-$Ga_2O_3$ vertical HJD.

## B. J-V & Reverse Breakdown Characteristics

The forward linear J-V characteristics of the $Cr_2O_3$/HVPE β-$Ga_2O_3$ and $NiO_x$/HVPE β-$Ga_2O_3$ HJDs on 100-μm dia. devices are shown in **Fig. 4(a)** by considering a 45°-angle current spreading[33,45]. Both types of HJDs exhibit comparable on-state current density in a range of 130-150 A/cm$^2$ at 5 V. The rectifying ratio of both types of HJDs are >$10^{10}$ with a reverse leakage of ~$10^{-8}$ A/cm$^2$ at -5 V and ideality factors of 1.72 and 1.63 for $Cr_2O_3$ and $NiO_x$ HJDs, respectively, as shown in **Fig. 4(b)**. The minimum differential specific on-resistance of the $Cr_2O_3$ and $NiO_x$-based HJDs are extracted to be 12.01 mΩ•cm$^2$ and 12.05 mΩ•cm$^2$, respectively, as shown in **Fig. 4(c)**. The reverse leakage and breakdown characteristics of the HJDs (submerged in FC-40 Fluorinert liquid) on HVPE β-$Ga_2O_3$ for 100-μm dia. devices are shown in **Fig. 4(d)**, showing a breakdown voltage range of 1.4-1.9 kV for $Cr_2O_3$/β-$Ga_2O_3$ HJDs and 1.5-2.3 kV for $NiO_x$/β-$Ga_2O_3$ HJDs with noise-floor reverse leakage current densities ($10^{-8}$~$10^{-6}$ A/cm$^2$, nA) at 80% of the devices' catastrophic breakdown voltage for both types of HJDs.

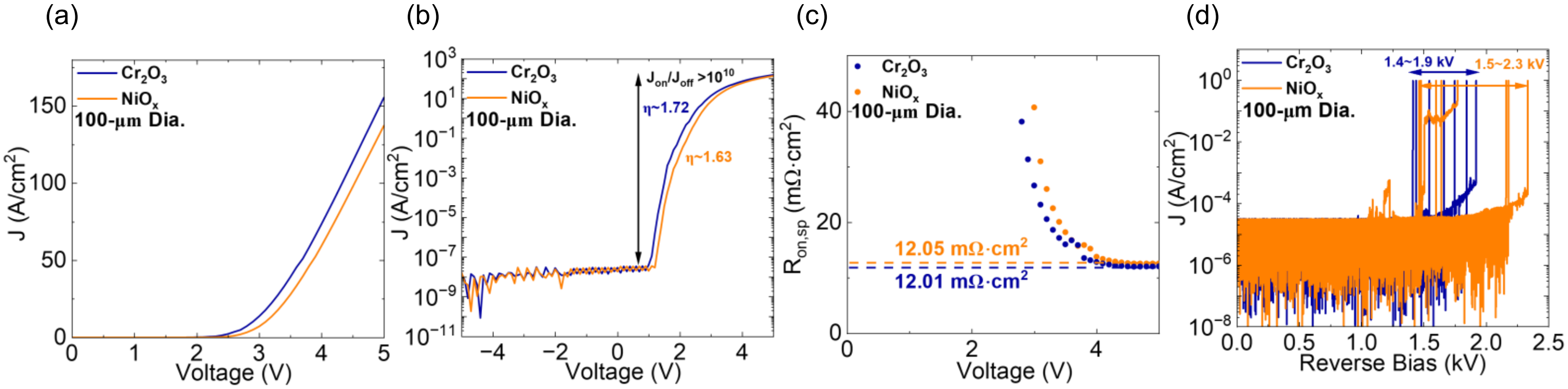

FIG.4. (a) Linear J-V characteristics of $Cr_2O_3$/β-$Ga_2O_3$ and $NiO_x$/HVPE β-$Ga_2O_3$ HJDs on 100-μm dia. device with a 45°-angle current spreading effect. (b) Semi-log scale J-V characteristics of $Cr_2O_3$/β-$Ga_2O_3$ and $NiO_x$/HVPE β-$Ga_2O_3$ HJD on 100-μm dia. device with a 45°-angle current spreading effect. (c) $R_{on,sp}$ vs. voltage of $Cr_2O_3$/β-$Ga_2O_3$ and $NiO_x$/HVPE β-$Ga_2O_3$ HJDs on 100-μm dia. device with a 45°-angle current spreading effect. (d) Reverse breakdown characteristics of $Cr_2O_3$/β-$Ga_2O_3$ and $NiO_x$/HVPE β-$Ga_2O_3$ HJD on 100-μm dia. devices.

## C. Electrical Stability under Ambient Conditions & Elevated Temperatures

The J-V characteristics of both as-fabricated $Cr_2O_3$/β-$Ga_2O_3$ and $NiO_x$/HVPE β-$Ga_2O_3$ HJDs were tracked as deposited, after 3 days, and after 10 days by exposing samples to the ambient. It was observed that the as-deposited current density of $Cr_2O_3$ HJD exhibited comparable value with that of $NiO_x$ HJD at 4 V, as shown in **Fig. 5(a)**. After 3 days of ambient exposure, the $Cr_2O_3$ HJD current density decreased slightly in the vicinity of 120 A/cm$^2$ at 4 V similar to what it was as deposited, but the $NiO_x$ HJD current density significantly dropped below 15 A/cm$^2$. After 10 days of ambient exposure, the $Cr_2O_3$ HJD remained around 120 A/cm$^2$ at 4 V while the $NiO_x$ HJD current density continued to decrease to < 1 A/cm$^2$, indicating electrical instability of $NiO_x$/HVPE β-$Ga_2O_3$ HJDs. To identify the cause of this degradation of the HJD's on-state degradation, $Cr_2O_3$ and $NiO_x$ layers were sputtered on sapphire with Ni/Au Ohmic contact for four-point Van Der Pauw measurements. It was observed that the sheet resistance ($R_{sh}$) of the sputtered $NiO_x$ on sapphire gradually increased to 50% of its original $R_{sh}$ value within 21 days, as shown in **Fig. 5(b)** while the $R_{sh}$ of $Cr_2O_3$ varied less than 10% of its original $R_{sh}$ value, implying enhanced electrical stability of $Cr_2O_3$ under ambient exposure compared to $NiO_x$. To further investigate the origin of the sheet resistance degradation in sputtered $NiO_x$ degradation, the $NiO_x$ on sapphire was exposed to different environments, and the sheet resistance variation was tracked after 72 hours exposure, as shown in **Fig. 5(c)**. A metal-organic chemical vapor deposition (MOCVD) reactor with high purity oxygen ($O_2$, 500 Torr) and nitrogen gas ($N_2$, 500 Torr) flow was used to expose the sample to $N_2$ and $O_2$. It was observed that the $R_{sh}$ of sputtered $NiO_x$ varied insignificantly within 72 hours when it was placed in pure $N_2$ and $O_2$, and the variation was comparable to the case where the reference sputtered $NiO_x$ on sapphire was sealed in vacuum post deposition. Upon submerging the $NiO_x$-on-sapphire sample in de-ionized water ($H_2O$), the $R_{sh}$ degraded ~10% within 72 hours, which was similar to the extent of degradation when the $NiO_x/Al_2O_3$ was exposed in ambient for the same amount of time. Therefore, it was qualitatively concluded that $H_2O$ is the agent that potentially causes the degradation in sputtered $NiO_x$.

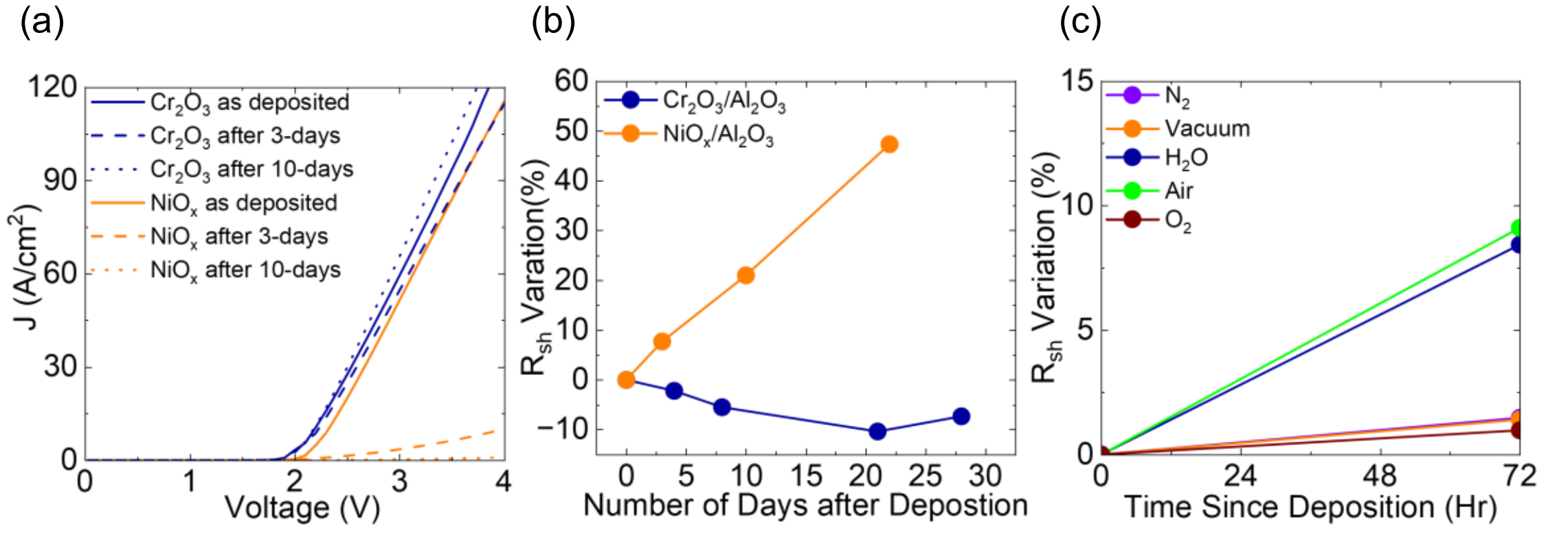

FIG.5. (a) Linear J-V characteristics of $Cr_2O_3$/β-$Ga_2O_3$ and $NiO_x$/HVPE β-$Ga_2O_3$ HJDs tracked as deposited, after 3-day, and after 10-day. (b). Sheet resistance variation of sputtered $Cr_2O_3$ and $NiO_x$ on reference sapphire wafers within 30-day. (c) Sheet resistance variation of sputtered $NiO_x$ under various exposure conditions within 72 hours.

Beyond J-V characterization of ambient exposed $Cr_2O_3$ HJDs, temperature-dependent J-V characteristics of the HJDs were measured to examine the diodes' thermal stability under elevated temperatures from 25 °C to 175 °C, as shown in **Fig. 6(a)**.

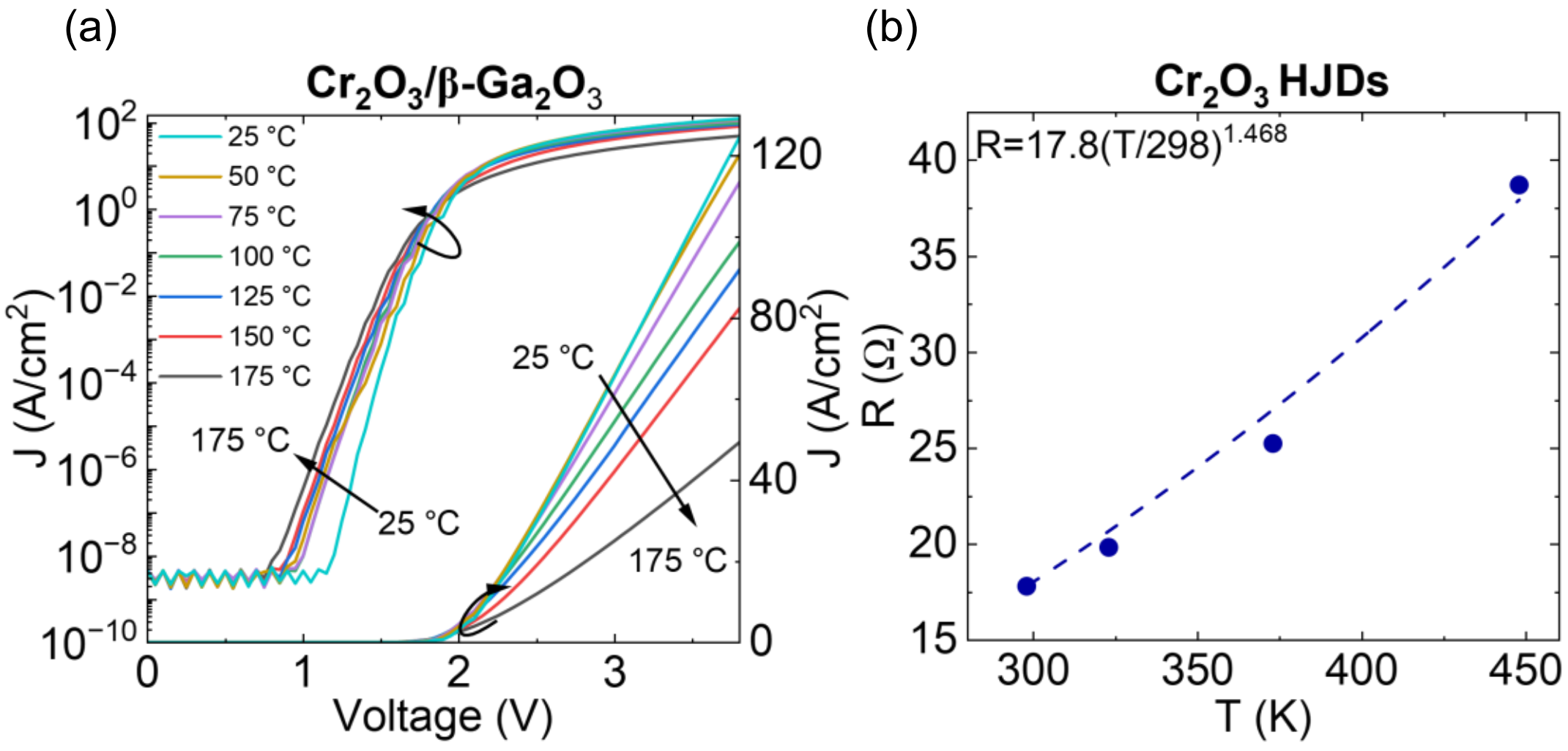

FIG.6. (a) Temperature-dependent J-V characteristics of $Cr_2O_3$/HVPE β-$Ga_2O_3$ HJDs from 25 °C to 175 °C in semi-log scale (left) and linear scale (right). (b) $Cr_2O_3$/HVPE β-$Ga_2O_3$ HJDs on-resistance vs. temperature power law fit.

The temperature-resistance dependent coefficient (α) was extracted from the power law fit of the on-resistance of the HJD as a function of temperatures, as shown in **Fig. 6(b)**. The extracted α of the $Cr_2O_3$/HVPE β-$Ga_2O_3$ HJD was lower than that of $NiO_x$/HVPE β-$Ga_2O_3$ HJD (α=1.56)[33] from previous literatures, indicating thermal stability[32] of sputtered $Cr_2O_3$ relative to $NiO_x$. Since the temperature-dependent sweep was carried out 10 days after the diode's ambient exposure, the $NiO_x$/HVPE β-$Ga_2O_3$ HJDs' on-state characteristics were heavily degraded, and no consistent variation trend was observed.

### D. Ternary (Ga-Ni-O, Ga-Cr-O) Compositional Phase Diagram

To further investigate the thermal instability in $NiO_x$/β-$Ga_2O_3$ heterojunction, the ternary compositional phase diagram of Ga-Ni-O and Ga-Cr-O systems generated by The Materials Project[46] database are compared in **Fig. 7(a) and 7(b)**, respectively, to examine any interfacial reaction hindering thermal stability of HJDs.

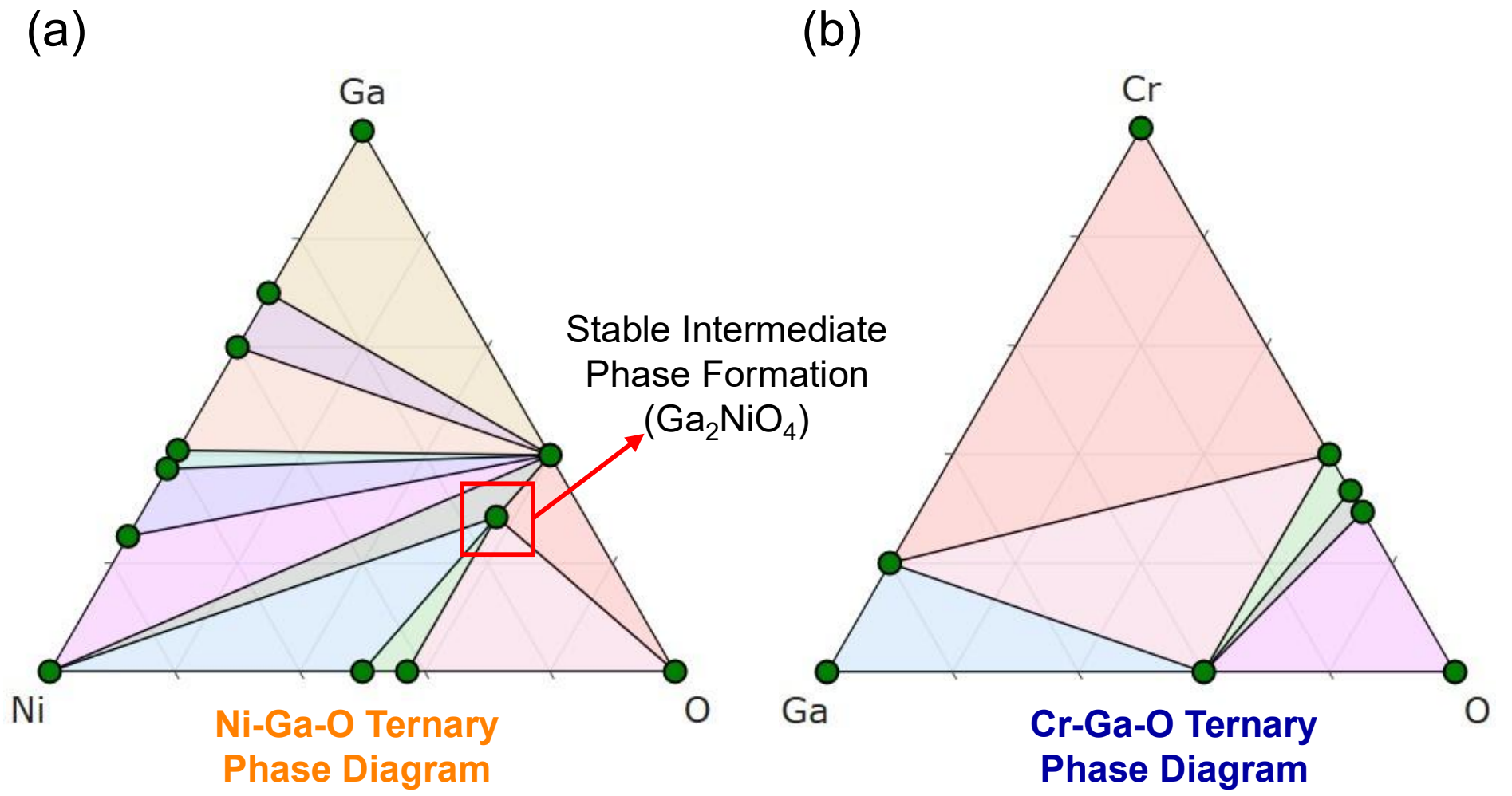

FIG.7. (a) Ternary phase diagram of Ga-Ni-O system. (b) Ternary phase diagram of Ga-Cr-O system.

Experimental observation of nickel gallate ($Ga_2NiO_4$)[25,26] for the $NiO_x$/β-$Ga_2O_3$ heterostructure system matches well with the stable phase predicted from the phase diagram. For the Ga-Cr-O system, there is no predicted stable intermediate phase formation at the $Cr_2O_3$/β-$Ga_2O_3$ heterojunction, indicating enhanced material system stability.

## V. Conclusion

In summary, this work explored the comparison of electrical performances of $NiO_x$/HVPE β-$Ga_2O_3$ and $Cr_2O_3$/HVPE β-$Ga_2O_3$ heterojunction diodes. It was discovered that the as-deposited $NiO_x$ and $Cr_2O_3$ HJDs exhibited similar electrical characteristics in terms reverse leakage current density, differential specific on-resistance, and breakdown voltages. A type-II band alignment between $Cr_2O_3$ and β-$Ga_2O_3$ was determined from first-principles calculations. The $NiO_x$ HJDs' forward conduction gradually degraded over extended period while the $Cr_2O_3$ HJDs' forward conduction remained nearly constant, and $H_2O$ is qualitatively identified as the agent that caused sheet resistance degradation in ambient exposed sputtered $NiO_x$. Temperature dependent J-V characteristics also revealed thermal stability in $Cr_2O_3$/HVPE β-$Ga_2O_3$ heterostructures relative to as-fabricated $Cr_2O_3$/HVPE β-$Ga_2O_3$ HJDs.

**ACKNOWLEDGMENTS**

The authors acknowledge funding from the U.S. Department of Energy (DOE) ARPA-E ULTRAFAST program (DE-AR0001824) and Coherent/II-VI Foundation Block Gift Program, and SUPREME, one of seven centers in JUMP 2.0, a Semiconductor Research Corporation program sponsored by the Defense Advanced Research Projects Agency. This research also used resources of the National Energy Research Scientific Computing Center, a DOE Office of Science User Facility supported by the Office of Science of the U.S. Department of Energy under Contract No. DE-AC02-05CH11231 using NERSC Award No. BES-ERCAP0028497. A portion of this work was performed at the UCSB Nanofabrication Facility, an open access laboratory.

**DATA AVAILABILITY**

The data that supports the findings of this study are available from the corresponding authors upon reasonable request.